\documentclass[twocolumn,tighten]{aastex702}

\usepackage{amsmath}

\begin{document}

\title{Excitation and Destruction of Small Polycyclic Aromatic Hydrocarbons by the Luminous Seyfert Nucleus of NGC\,4151}

\author[0009-0001-6065-0414]{Grant P. Donnelly}
\affiliation{Ritter Astrophysical Research Center, University of Toledo, Toledo, OH 43606, USA}
\email[show]{Grant.Donnelly@rockets.utoledo.edu}

\author[0000-0003-1545-5078]{J.-D. T. Smith}
\affiliation{Ritter Astrophysical Research Center, University of Toledo, Toledo, OH 43606, USA}
\email{JD.Smith@utoledo.edu}

\author[0000-0002-6353-1111]{Sebastian F. H{\"o}nig}
\affiliation{School of Physics and Astronomy, University of Southampton, Southampton, SO17 1BJ, United Kingdom}
\email{S.Hoenig@soton.ac.uk}

\author[0000-0002-5782-9093]{Daniel~A.~Dale}
\affiliation{Department of Physics and Astronomy, University of Wyoming, Laramie, WY 82071, USA}
\email{ddale@uwyo.edu}

\author[0000-0003-0699-6083]{Tanio D\'iaz-Santos}
\affiliation{School of Sciences, European University Cyprus, Diogenes street, Engomi, 1516 Nicosia, Cyprus}
\affiliation{Institute of Astrophysics, Foundation for Research and Technology-Hellas (FORTH), Heraklion, 70013, Greece}
\email{tanio@ia.forth.gr}

\author[0000-0002-0846-936X]{B.~T.~Draine}
\affiliation{Dept.\ of Astrophysical Sciences, Princeton University, Princeton, NJ 08544, USA}
\email{draine@astro.princeton.edu}

\author[0000-0001-7449-4638]{Brandon~S.~Hensley}
\affiliation{Jet Propulsion Laboratory, California Institute of Technology, 4800 Oak Grove Drive, Pasadena, CA 91109, USA}
\email{brandon.s.hensley@jpl.nasa.gov}

\author[0000-0001-8490-6632]{Thomas S.-Y. Lai}
\affil{IPAC, California Institute of Technology, 1200 E. California Blvd., Pasadena, CA 91125, USA}
\email{ThomasLai.astro@gmail.com}

\author[0000-0002-4378-8534]{Karin M. Sandstrom}
\affil{Department of Astronomy \& Astrophysics, University of California, San Diego, 9500 Gilman Drive, La Jolla, CA 92093, USA}
\email{kmsandstrom@ucsd.edu}

\author[0000-0002-0710-3729]{Andrey Vayner}
\affiliation{Florida Gulf Coast University, 10501 FGCU Blvd. South, Fort Myers, FL 33965, USA}
\email{anvayner@fgcu.edu}

\begin{abstract}
	
We present a spatially-resolved detection of $3.3\,\mathrm{\mu m}$ emission from polycyclic aromatic hydrocarbons (PAHs) within the circumnuclear ${{\sim}}200\,\mathrm{pc}$ of the archetypal Seyfert galaxy NGC\,4151, using new JWST NIRSpec/IFU spectroscopy. PAHs are detected down to the innermost $50\,\mathrm{pc}$, though it is likely that PAHs also exist at smaller radii, as PAH detections closer than this seem to be limited by continuum dilution. This is contrary to expectations that PAH grains, especially the smallest grains that carry the $3.3\,\mathrm{\mu m}$ feature, could not survive near an active galactic nucleus (AGN). Furthermore, the surface brightness of PAH emission increases as radius from the nucleus decreases. We interpret this as evidence that the AGN supplies energy for PAH excitation by driving ultraviolet emission in nearby H and $\mathrm{H_2}$ gas. We do not detect PAH emission within the narrow line region (NLR) that hosts highly ionized and outflowing gas, suggesting that the AGN may also destroy PAHs in these regions, depending on geometry. Together, these results show that an AGN has a complex, multifaceted role in the life cycle of PAHs and in PAH excitation.

\end{abstract}


\section{Introduction}
\setcounter{footnote}{0} 

Polycyclic aromatic hydrocarbons (PAHs) are among the smallest forms of dust in the interstellar medium (ISM). Upon absorbing light from the surrounding interstellar radiation field (ISRF), PAHs re-emit this energy as a family of broad spectral features in the mid-infrared (MIR) \citep{leger_puget_1984, allamandola_1985}. As the efficiency at which a PAH grain emits a particular spectral feature is determined by factors like grain size, charge, and temperature \citep{draine_21}, PAH luminosity and feature ratios are sensitive to conditions in the ISM. PAHs track the star-formation rate (SFR) \citep{peeters_2004, shipley_2016, lai_2020}, the gas-phase metallicity \citep{whitcomb_2024} and the molecular gas content \citep{chown_2025}, and serve as a probe for characteristics of the PAH population itself like the grain size distribution (GSD) and ionization \citep{draine_21}. A long-standing problem concerns PAHs in galaxies hosting an active galactic nucleus (AGN): spectra from PAHs are markedly different in AGN galaxies compared with non-AGN galaxies. What impact does an AGN have on the physical characteristics and emission properties of PAH grains? 

Spectra of the central regions of AGN galaxies consistently show reduced or no PAH emission \citep[e.g.,][]{aitken_roche_1985, moorwood_1986, tran_2001, siebenmorgen_2004, desai_2007, smith_2007, honig_2010, lamassa_2012}. This is often interpreted as evidence that an AGN destroys PAHs via photodestruction \citep{aitken_roche_1985, desert_1988, voit_1992, siebenmorgen_2004, monfredini_2019} or through mechanical interactions such as shocks \citep{micelotta_2010a, micelotta_2010b, diamond-stanic_2010, zhang_2022}, although some of this effect may be attributable to dilution from the overwhelmingly bright AGN continuum \citep{lutz_1998, clavel_2000, desai_2007, diaz-santos_2010, alonso-herrero_2014, xie_ho_2022}. Still, in spatially-resolved studies of PAHs in AGN hosts using Spitzer and now JWST, PAH features are almost always seen in the spectra of the resolved extended regions of AGN galaxies \citep[e.g.,][]{smith_2007, diamond-stanic_2010, lai_2022, lai_2023, garcia-bernete_2024, donnelly_2024} and even quasar hosts \citep{neugebauer_1979, schweitzer_2006, xie_ho_2022, ramos-almeida_2023, ramos_almeida_2025}. Likewise, there are many examples of PAHs being detected within the central few hundred parsecs of AGN using ground-based spectroscopy \citep{diaz-santos_2008, honig_2010, gonzalez-martin_2013, ramos-almeida_2014, esquej_2014, jensen_2017, esparza-arredondo_2018, alonso-herrero_2020}, complicating the picture that AGN are a destructive force on PAH molecules. 

An additional complexity is that the profiles of PAH feature ratios signal an influence from an AGN; PAH populations closer to the nucleus apparently lack smaller grains, leading to the idea that AGN may selectively destroy the smallest PAH grains, while other larger grains are able to survive \citep{smith_2007, odowd_2009, sales_2010, garcia-bernete_2022a, zhang_2022, lai_2022, garcia-bernete_2022b, lai_2023, garcia-bernete_2024, zhang_2024}. If this is confirmed, the presence of an AGN would need to be considered alongside the other processes that impact the life cycle of PAHs. Likewise, such a connection between PAHs and an AGN would become an important mechanism to include within the growing number of galaxy simulations that take PAHs into account \citep[e.g.,][]{narayanan_2023, narayanan_2026}. Finally, there has even been some evidence that an AGN can supply a portion of the photons for PAH excitation \citep{howell_2007, jensen_2017}, adding another important impact of an AGN on PAHs.

The $3.3\,\mathrm{\mu m}$ PAH feature is produced primarily by the smallest PAH grains, as only the smallest grains with 20--50 C atoms (which have the smallest heat capacities) can be briefly heated to the high temperatures required to efficiently emit at these relatively short wavelengths \citep{draine_21}. Thus, this feature is the most sensitive tracer of PAH size when compared to the strengths of other features. This quality perhaps makes the $3.3\,\mathrm{\mu m}$ feature the most important for addressing the impact of an AGN on PAHs and the proposed small-grain destruction effect. Despite the progress made towards understanding PAHs in AGN hosts by Spitzer, the IRS spectrometer only could observe at $\lambda>5.2\,\mathrm{\mu m}$, so comparisons with longer-wavelength PAH features have been limited. Before JWST, the $3.3\,\mathrm{\mu m}$ PAH has been detected in AGN galaxies using space-based facilities like ISO \citep[e.g.,][]{clavel_2000} and AKARI \citep[e.g.,][]{ichikawa_2014, inami_2018, lai_2020, garcia-bernete_2022a}. However, only JWST NIRSpec possesses the sensitivity and spatial resolution to search for $3.3\,\mathrm{\mu m}$ PAH emission within the central ${{\sim}}100\mathrm{s\,pc}$ near an AGN free from contamination by the nuclear emission. 

Spatially-resolved PAH maps of the inner few hundred parsecs of a luminous, nearby AGN have the power to probe an environment where the impact of an AGN on PAHs is most evident. Can small PAHs survive near an AGN? If so, where? Can an AGN directly excite PAHs? In this work, we address these questions using new NIRSpec IFU spectroscopy of the center of NGC\,4151, one of the most nearby and best-studied Seyfert galaxies. In Section~\ref{sec:data_methods}, we describe the NIRSpec observation and data reduction, and we describe how spectra are extracted and how emission features are fit. In Section~\ref{sec:results} we describe the results of these analyses, and in Section~\ref{sec:discussion} we discuss the implications of these results. Finally, this work is summarized in Section~\ref{sec:summary}.

\section{Data and Methods} \label{sec:data_methods}

\subsection{The Target: NGC\,4151}

The circumnuclear environment of NGC\,4151 presents a unique opportunity to build upon the growing body of work on PAHs in AGN hosts. It is close-by \citep[Cepheid distance $D=15.8\,\mathrm{Mpc}$,][]{distance}, relatively face-on \citep[$i=21^{\circ}$,][]{inclination}, and has one of the brightest AGN of any nearby galaxy \citep[mean $L_{\mathrm{2-10, keV}}=10^{42.73}\,\mathrm{erg\,s^{-1}}$, adjusted for $D=15.8\,\mathrm{Mpc}$,][]{Lx}, though it is highly variable. As one of the original Seyfert galaxies \citep{seyfert_43}, this target benefits from having an extremely well-studied nuclear environment. NGC\,4151 contains a bi-conical narrow line region (NLR) of highly ionized material with a $\mathrm{PA=60^{\circ}}$ and an inclination into the plane of the sky of $i=45^\circ$, where the SW cone is closer to us, and with an outer half-opening angle of $33^\circ$ \citep[][]{crenshaw_2000, das_2005, fischer_2014}. These cones consist of outflowing gas with line-of-sight velocities of more than $\pm 600\,\mathrm{km\,s^{-1}}$, with the NE part receding and the SW part approaching \citep{das_2005, storchi-bergmann_2010}. This orientation permits study of how energy emitted by the nucleus impacts the host galaxy itself, including PAHs; for example, \cite{mundell_2003} detected a radio jet originating from the nucleus that is slightly misaligned with the NLR cones, but which is likely interacting with NLR material \citep{storchi-bergmann_2010}. 

The nuclear environment is also dusty; using reverberation mapping, \cite{lyu_2021} detected an extended dust component within the central few hundred parsecs, suggesting that there should be a component of warm polar dust within the ionization cones, and also cold dust in the perpendicular region that is shielded by the torus. Based on CO (1-0) \citep{dumas_2010} and Hubble $V-I$ \citep{asif_2005} observations, dust lanes approach the nucleus from the NW and SE \citep{wang_2011}. Indeed, this was one of the first AGN galaxies where PAH $3.3\,\mathrm{\mu m}$ was detected in ${\sim}1\,$kpc-scale spectra \citep{cutri_1980}. These factors make NGC\,4151 a laboratory for looking at the effects of various possible physical processes acting on PAHs at relatively small scales.

\subsection{Observations}

\begin{figure*}
	\centering
	\includegraphics[width=0.99\textwidth]{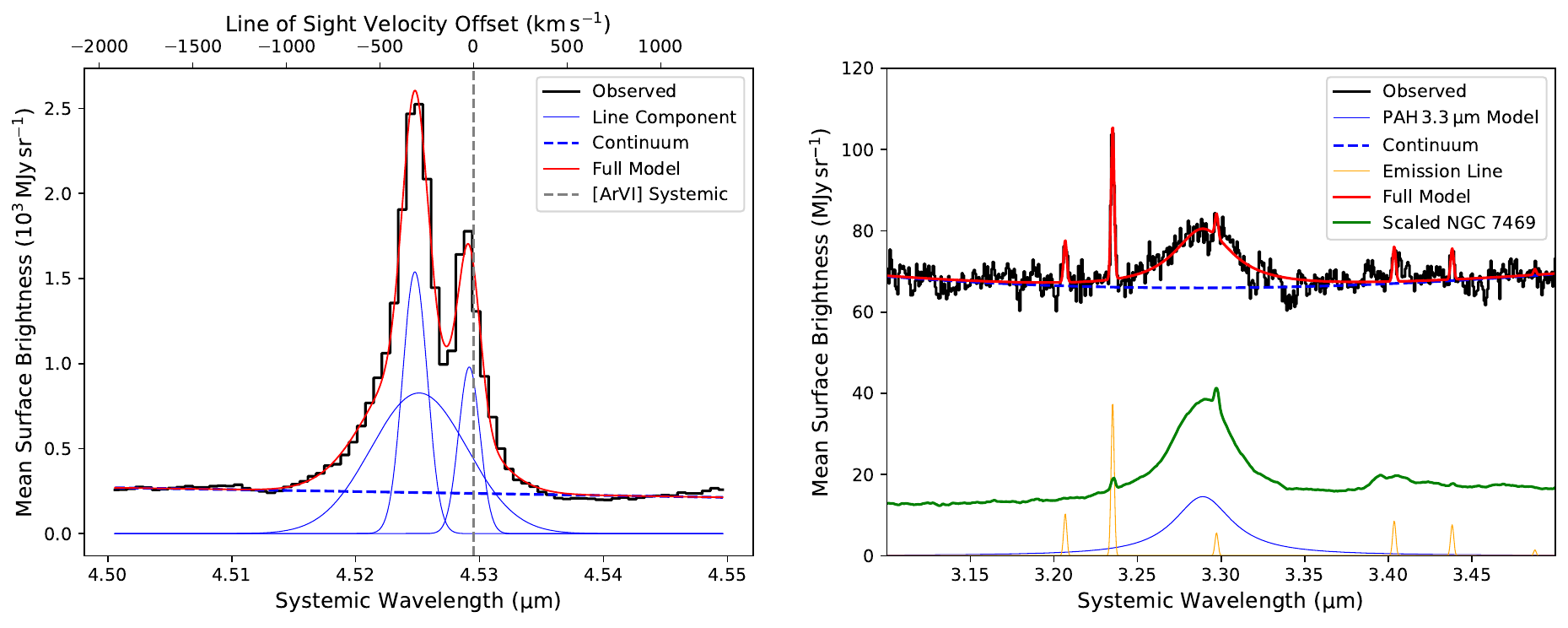}
	\caption{
	Examples of model fits to different components of spaxel spectra. Left: the [ArVI] line for this example $5\times5$ spaxel region is fit with three Gaussian components, which is the maximum that we allow, though in other regions this line is fit with zero, one or two. The continuum is assumed to be linear. Right: the fit for the PAH $3.3\,\mathrm{\mu m}$ feature for each region is modeled as a Drude profile, and the continuum is modeled with a third-degree polynomial. Several emission lines (orange) are fit as well, although many of these are not prominent in this example. We also show the ${\sim}3.3\,\mathrm{\mu m}$ spectrum from the circumnuclear starburst ring of NGC\,7469 for comparison in green; this spectrum may be considered a typical PAH emission profile. In the central regions of NGC\,4151 like in this example, the PAH $3.3\,\mathrm{\mu m}$ feature has a smaller equivalent width.
	}
	\label{fig:spaxel_fits}
\end{figure*}

This work is focused on NIRSpec IFU spectroscopy centered on the nucleus of NGC\,4151, as part of the Cycle 4 JWST program 8660 (P.I. Donnelly). The observations were performed on May 11, 2026 with a total exposure time of 1460.2\,s, using a 4-point dither, together with \texttt{leakcal} integrations that match the exposure time and dither pattern on source. We used the high-resolution G395H/F290LP grating/filter configuration, which covers $\lambda=2.87–5.27\,\mathrm{\mu m}$ (or $\lambda_{\mathrm{rest}}=2.86–5.25\,\mathrm{\mu m}$ when corrected for $z=0.003326$) at $R{\sim}2700$. The high spectral resolution aids in resolving the various kinematic components of emission lines in the NLR, although the chip-gap prevents observations between about $3.98\,\mathrm{\mu m}$ and $4.20\,\mathrm{\mu m}$ ($3.97\,\mathrm{\mu m}$ and $4.19\,\mathrm{\mu m}$ when redshift-corrected), depending on the cube spaxel. At $D=15.8\,\mathrm{Mpc}$, the $3\,\arcsec$ edge-length of the NIRSpec IFU FOV covers about $230\,\mathrm{pc}$.

We downloaded the \texttt{uncal} observation files from the MAST archive and reduced these into data cubes using the JWST NIRSpec pipeline example notebook with version \texttt{v2.0.0} and reference file \texttt{jwst\_1364.pmap}. Subtraction of the \texttt{leakcal} frames was done as part of the reduction. For the most part, we used the default pipeline settings, but we used the \texttt{adaptive\_trace\_model} pipeline option to reduce the spectral ringing.

The initial data cube requires that the point spread function (PSF) of the nuclear source be subtracted. We describe our PSF subtraction method in detail in Appendix~\ref{appx:PSFsub}. Two additional pipeline settings were different from the defaults in order to facilitate the PSF subtraction. First, we set the spaxel size to $0\farcs05$ (the default is $0\farcs1$) for better PSF sampling and centroid fitting. For NGC\,4151, this gives ${\sim3.8\,\mathrm{pc}}$ per spaxel. Second, the cube was reduced in the \texttt{ifualign} internal coordinate system so that the PSF orientation on the cube is independent of the telescope roll angle. All images in this work, for the purposes of both analysis and presentation, are kept in this orientation to avoid unnecessary interpolation. We note also that some of the spaxels nearest to the AGN have been saturated, and there is a horizontal ``bar" artifact resulting from central saturation; this artifact has been noted in other observations \citep{boker_2022} and we mask the bar for this work. All of the following analysis is done on the PSF-subtracted data.

\subsection{Spectral Extraction and Fitting}

In the following subsections, we describe our processes for fitting the various spectral features present in the NIRSpec IFU spectroscopy, and how we construct spatially-resolved integrated intensity maps of these features.

\subsubsection{Emission Line Fitting} \label{sec:line_fitting}

The spectra within the central ${\sim}\mathrm{200\,pc}$ of NGC\,4151 exhibits a variety of emission lines including H and He recombination lines, $\mathrm{H_2}$ molecular rovibrational lines, and a collection of fine-structure lines. We choose the coronal line [ArVI] to trace the extent of the NLR, given its high IP (91.29\,eV) and that it is one the brightest lines that we observe within the NLR in the NIRSpec spectra. In order to explore the relationship between $\mathrm{H_2}$ abundance/excitation and PAH survival/excitation, we also fit any detected $\mathrm{H_2}$ rovibrational lines. 

For each spaxel $i,j$, we extract the mean spectrum of a $5\times5$ spaxel kernel centered at $i,j$. This smoothing promotes a better signal-to-noise ratio (SNR) at the cost of some spatial detail. For each emission line, we isolate the section of the spectrum within a wavelength window chosen to include a significant stretch of continuum compared with the line profile width while excluding any other lines; this typically amounts to a ${\sim}0.045\,\mathrm{\mu m}$ window centered on the systemic wavelength of the line. 

The extracted emission line profiles in the NLR exhibit complex kinematics, evident from different degrees of broadening and often with multiple narrow peaks. Thus, each redshift-corrected line profile is fit with a model that is the sum of a linear function for the continuum, and between zero and three Gaussian profiles that correspond to the line emission. We demand that the amplitudes of each profile are non-negative, and the central wavelengths and widths of each Gaussian profile are allowed to float within sensible boundaries; the central wavelength can be offset from rest by no more than 1200\,$\mathrm{km\,s^{-1}}$ for [ArVI] and 500\,$\mathrm{km\,s^{-1}}$ for $\mathrm{H_2}$ lines. We limit the full-width at half-maximum (FWHM) for each Gaussian component to no lower than 80\,$\mathrm{km\,s^{-1}}$ for all lines, and no higher than 235\,$\mathrm{km\,s^{-1}}$ for $\mathrm{H_2}$ and 700\,$\mathrm{km\,s^{-1}}$ for [ArVI]. Each spectrum is fit four times independently with increasing numbers of Gaussian profiles (starting with zero), then a statistical test is applied to determine which model is most appropriate. In general, we wish to select the best fit that uses the fewest free parameters (i.e., Gaussian profiles); in this work, we are only interested in describing the line profiles so that the true intensity is obtained, and having fewer Gaussian components reduces the risk of erroneously fitting line components to noise. Thus, we apply the Akaike Information Criterion \citep[AIC,][]{akaike_1974, thorngren_2026}, which can be used to compare the goodness-of-fit between models while penalizing models with more free parameters. 

\begin{figure*}
	\centering
	\includegraphics[width=0.99\textwidth]{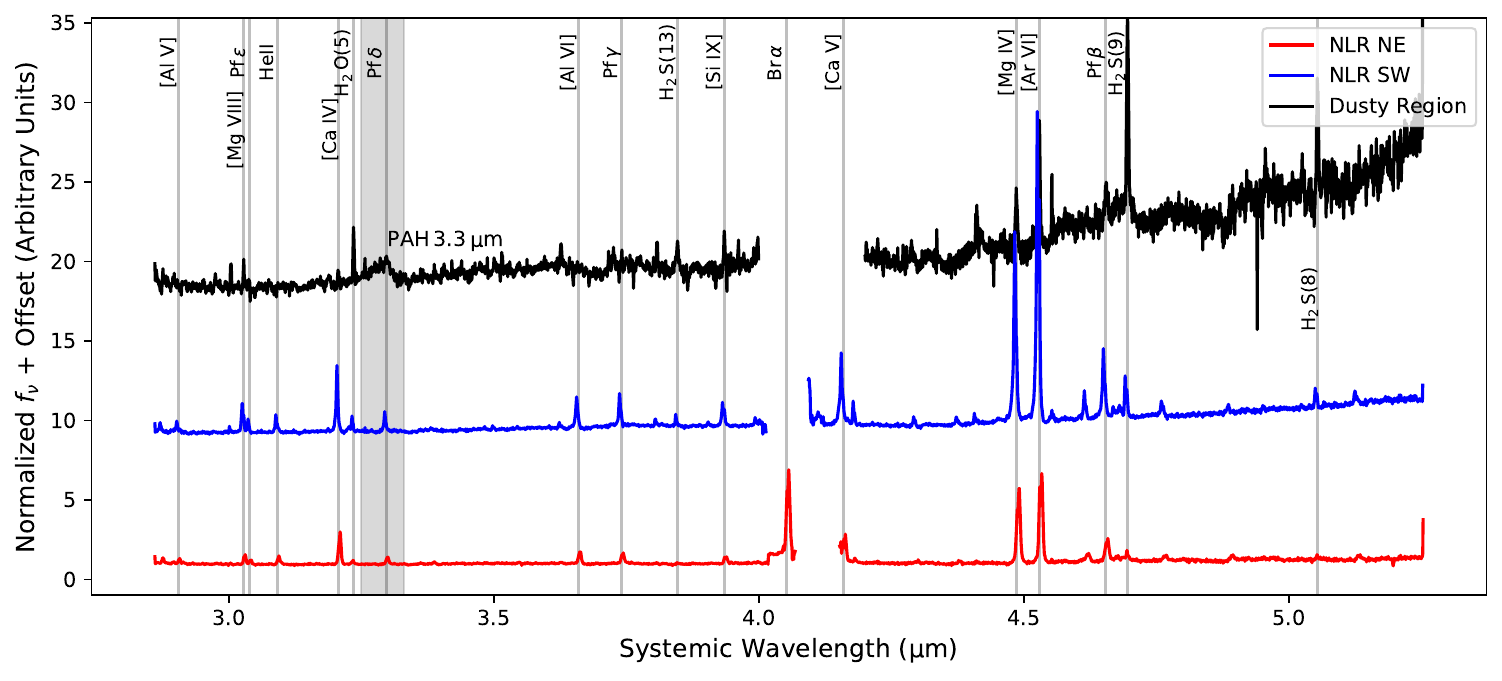}
	\caption{
		The red and blue spectra are taken from $r=0\farcs2$ circular apertures in the NE and SW parts of the NLR, respectively, with the brightest emission lines labeled. These two spectra have a common normalization at the $3.5\,\mathrm{\mu m}$ continuum, with offsets applied for clarity. The ${\sim}4.1\,\mathrm{\mu m}$ gap is slice-dependent on the IFU, thus lines like [CaV] and $\mathrm{Br}\,\alpha$ can only appear in certain parts of the cube, even if they are otherwise detectable. The gray vertical lines indicate the systemic wavelength of each emission line; the bulk motion of material in the NE is redshifted, and material in the SW is blueshifted. However, many of the lines exhibit multiple kinematic components (e.g., Figure \ref{fig:spaxel_fits}). The black spectrum is taken from a dusty, relatively PAH-bright region (see Figure~\ref{fig:PAH_map}) and serves as an example spectrum with a $3.3\,\mathrm{\mu m}$ PAH feature. Note that the PAH feature is not present in the NLR spectra.
	}
	\label{fig:outflow_spectra}
\end{figure*}

In our case, for each spectrum, the AIC is obtained for all four model fits and the model with the lowest AIC is selected. Thus, an additional Gaussian component is only allowed if it would significantly improve the fit. This procedure is mostly relevant for fitting [ArVI], which is brightest in the NLR where outflows are present. An example of a fit for the [ArVI] line in a three-component region is given in Figure~\ref{fig:spaxel_fits}. Though the same procedure is used for fitting the $\mathrm{H_2}$ lines (which are typically faint in NLR spaxels), these usually have only one or two kinematic components with a modest velocity offset from rest due to normal galaxy rotation.

We generate flux maps from the spectral fits. To do this, we first compute the mean line flux within the extraction kernel as the sum of the integrals of each Gaussian component, independent of the continuum. This integrated quantity is the mean flux per unit solid angle within the $5\times5$ spaxel kernel centered on $i,j$. Finally, all integrations that pass the cut are reassembled into a two-dimensional image where each pixel $i,j$ has the mean intensity value from its $5\times5$ spaxel extraction.

\subsubsection{PAH $3.3\,\mathrm{\mu m}$ Fitting}

Spectral extraction for PAH fitting is performed in a similar way as for the line extractions, using the same $5\times5$ spaxel moving kernel. For these extractions, we fit the spectrum between $\lambda=3.1-3.5\,\mathrm{\mu m}$. Spectral PAH features take the shape of Drude profiles \citep[e.g.,][]{smith_2007}, which are broad features with more power in their wings compared with a Gaussian of the same FWHM. We adopt the PAH $3.3\,\mathrm{\mu m}$ profile parameters of \cite{draine_2025}; a single Drude profile with $\lambda_{\mathrm{cen}}=3.289\,\mathrm{\mu m}$, and $\mathrm{FWHM}=0.0423\,\mathrm{\mu m}$. We do not fit the $3.4\,\mathrm{\mu m}$ feature, attributed to aliphatic grains (e.g., green spectrum in Figure~\ref{fig:spaxel_fits} right) as it is not apparent in the NGC\,4151 data even for regions of high PAH SNR.

Our model to fit the PAH emission then consists of a linear combination of one Drude profile as described above, which represents the PAH feature, a third-degree polynomial for the continuum, and single-component Gaussian profiles for various emission lines. For the continuum, we find that such a polynomial is necessary as there is often significant curvature present in the continuum over this wider wavelength range when compared with the line fits. We fit six emission lines that can appear in these spectra with enough intensity to otherwise affect the Drude and continuum fits: including [CaIV] $\lambda3.2067\,\mathrm{\mu m}$, $\mathrm{H_2\,O(5)}$ $\lambda3.2350\,\mathrm{\mu m}$, $\mathrm{Pf\,\delta}$ $\lambda3.2970\,\mathrm{\mu m}$, and lines at 3.4045, 3.4380, and 3.4860\,$\mathrm{\mu m}$. In Figure~\ref{fig:spaxel_fits}, we show an example fit for a particularly high (relative to other spaxels) SNR PAH detection.

To avoid erroneously fitting a PAH feature to spectra with residual ringing artifacts, we exclude extractions where the spectral standard deviation within a defined continuum window ($3.2>\lambda>3.4\,\mathrm{\mu m}$) is greater than 10\% of the median value of the spectrum. We place an additional cut at $\mathrm{SNR}\geq3$ for PAH feature fits to determine spectra of good enough quality to contribute to a pixel of a given map. The PAH map is then generated from the spectral fits in the same way as the integrated line intensity maps.

\section{Results} \label{sec:results}

\subsection{Representative Spectra from Outflow and Dusty Regions}

We begin by showing spectra from three representative apertures: two regions in the NLR along the axis of the outflow/bicones, with one in the northeast (NE) and one in the southwest (SW) ($\mathrm{PA}=60^\circ$ and $240^\circ$ from the nucleus, respectively), and one aperture in a dusty region outside of the NLR selected for its high PAH SNR ($\mathrm{PA}=-30^\circ$). Each region is equidistant from the nucleus at $75\,\mathrm{pc}$, and is a circle with $r=0\farcs2$ ($r=15.3\,\mathrm{pc}$). These representative spectra are shown in Figure~\ref{fig:outflow_spectra}. We adopt the absolute J2000 coordinates for the AGN at $\alpha=\mathrm{12^{h}10^{m}32^{s}.5758}, \delta=\mathrm{39^{\circ}24^{\prime}}21^{\prime\prime}.060$ from \cite{AGN_coords}, which were determined from VLBA observations.

Both of the NE and SW NLR spectra include strong coronal lines with high IPs that are associated with AGN activity (e.g., [Mg\,IV], [ArVI], [Ca\,IV]), including the [Si\,IX] line with $\mathrm{IP=351.28\,eV}$, as well as H recombination lines. The NLR spectra also show $\mathrm{H_2}$ rotational lines, although these are rather weak in comparison. The spectrum extracted for the dusty region shows that it is a much less extreme environment compared with the NLR in terms of the ionization state and gas velocities; the coronal lines are present but are much weaker, and the H recombination lines are weak or absent. However, the rovibrational $\mathrm{H_2}$ lines are strong in the dusty region.  A more quantitative description of spatially-resolved PAH detections in NGC\,4151 follows below.

\subsection{Spatial Distribution of PAH Intensity} \label{sec:PAH_maps}

\begin{figure}
	\centering
	\includegraphics[width=0.49\textwidth]{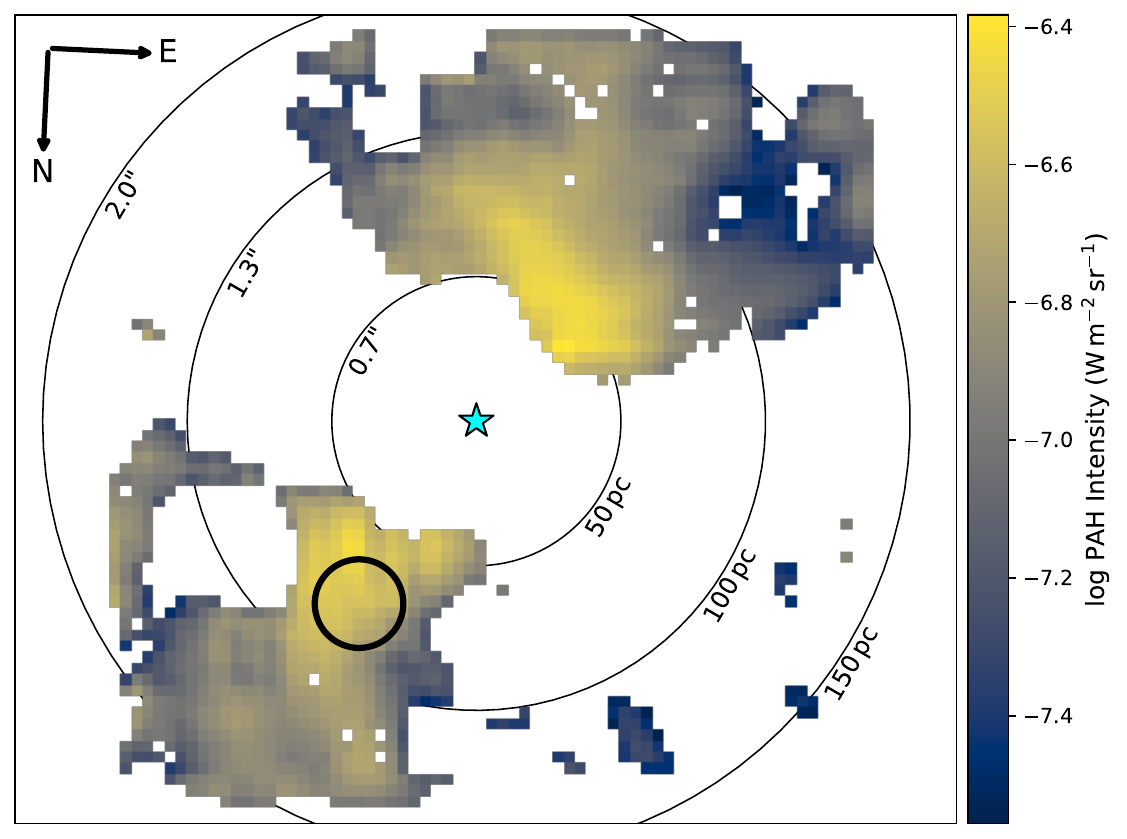}
	\caption{
    Spatial distribution of detected PAH $3.3\,\mathrm{\mu m}$ within the central ${\sim}200\,\mathrm{pc}$ of NGC\,4151. The cyan star is the location of the AGN, and concentric circles give both the spatial and angular distances from the nucleus. The black circle in the NW corner indicates the dusty region for the spectral extraction shown in Figure~\ref{fig:outflow_spectra}. PAH emission is detected down to $\lesssim50\,\mathrm{pc}$ from the nucleus, and it increases in intensity as radius decreases.
	}
	\label{fig:PAH_map}
\end{figure}

In Figure~\ref{fig:PAH_map}, we show the spatial distribution of PAH $3.3\,\mathrm{\mu m}$ intensity. The PAH-emitting regions are situated to the northwest (NW) and southeast (SE) of the nucleus, with morphology that may be tracing the innermost parts of dusty spiral arms winding towards the center. Indeed, the NW and SE are where spiral arms detected via CO (1-0) at $r{\sim}1\,\mathrm{kpc}$ \citep{dumas_2010} would likely intercept the nucleus. PAHs are detected within as close as $\lesssim50\,\mathrm{pc}$ from the center, and there is an indication from this map that the PAH intensity increases as distance to the nucleus decreases (discussed further in Section~\ref{sec:excitation}). This is contrary to many expectations that no PAHs could survive within such an environment, especially not the smallest grains which are the primary carriers of the PAH $3.3\,\mathrm{\mu m}$ feature. Notably, we do not detect PAH emission along the direction of the NLR to the NE and SW of the nucleus.

As PAH spectral features are broad, and a significant portion of their power is in their shallow wings, PAH features are particularly susceptible to continuum dilution. This is often the case for AGN galaxies, as an AGN is especially effective at producing a strong continuum at shorter wavelengths from thermal emission of a hot, dusty, circumnuclear torus \citep{alonso-herrero_2001, nenkova_2002}. Indeed, the measured PAH equivalent width (EQW) is often used as an indicator for the presence of an AGN, with AGN galaxies typically having lower PAH EQWs, due to grain destruction, continuum dilution, or both \citep{genzel_1998, lutz_1998, laurent_2000, desai_2007, dale_2009, petric_2011, stierwalt_2014, rich_2023}. The PAH $3.3\,\mathrm{\mu m}$ EQW near the NGC\,4151 nucleus is very low, where the EQW for most spaxels is between 3 and 15\,nm; starburst galaxies will typically have PAH $3.3\,\mathrm{\mu m}$ EQW ${\sim}100\,\mathrm{nm}$, whereas galaxies with an AGN often have EQW $<40\,\mathrm{nm}$ \citep{ichikawa_2014}. We discuss the extent to which continuum dilution and/or an actual absence of PAH grains may explain the lack of PAH emission at smaller radii and in the NLR in Section~\ref{sec:discussion}.

\subsection{Separation of PAHs and Ionized Gas} \label{sec:results_distros}

\begin{figure}
	\centering
	\includegraphics[width=0.49\textwidth]{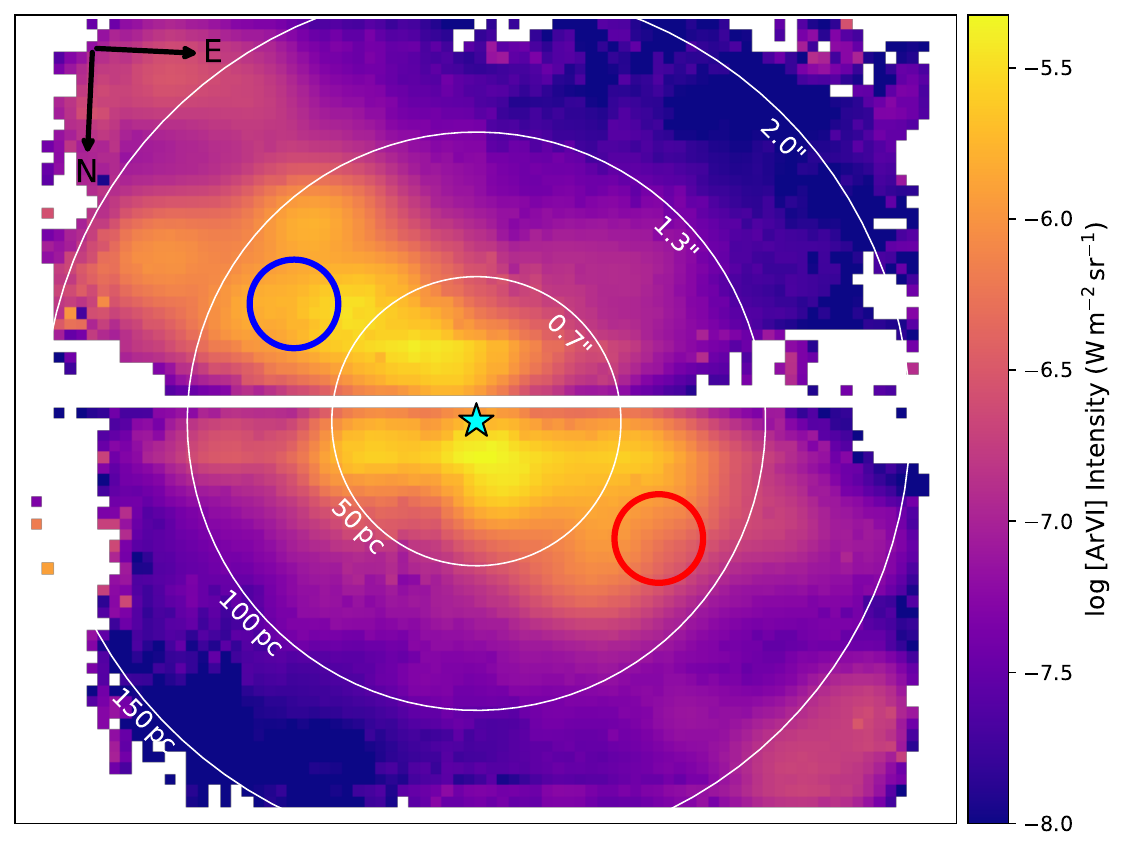}
	\caption{
    Similar to Figure~\ref{fig:PAH_map}, but for the integrated intensity of the [ArVI] line. The blue and red circles show the regions for the NLR spectral extractions shown in Figure~\ref{fig:outflow_spectra}. The [ArVI] emission traces the NLR that extends towards the NE and SW of the nucleus.
	}
	\label{fig:ArVI_map}
\end{figure}

\begin{figure*}
	\centering
	\includegraphics[width=1\textwidth]{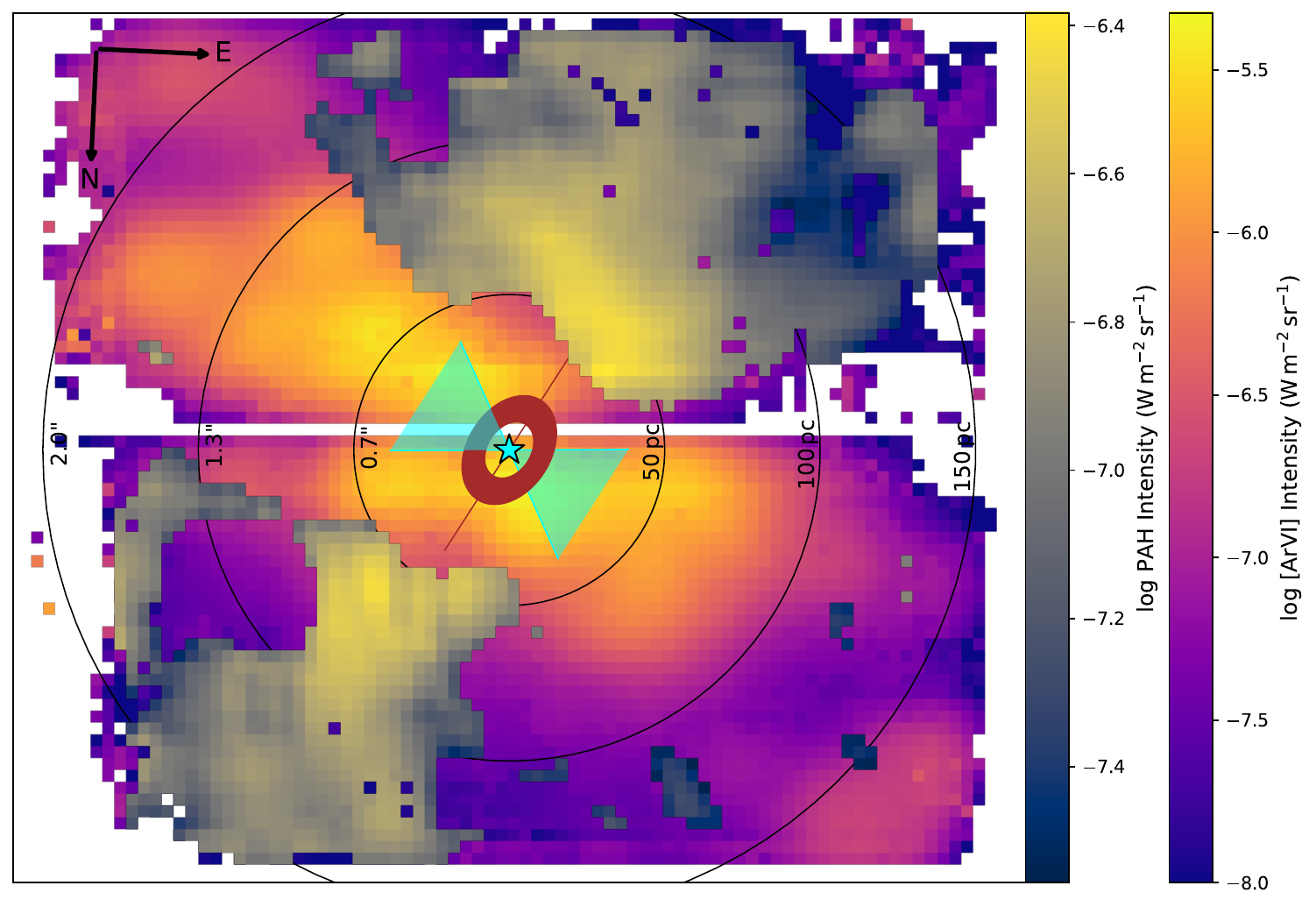}
	\caption{
	Two separate distributions are shown, combining together the PAH and [ArVI] maps in Figures~\ref{fig:PAH_map} and \ref{fig:ArVI_map}, respectively. The cyan cones show the direction and opening angle of the bi-conical outflow \citep{das_2005}. The brown ellipse represents the putative circumnuclear torus; it has been rotated around the brown line axis to show the inclination of the nuclear system, whereas a circle would indicate no inclination. These structures are not to scale, but are intended only to show the orientation and location of the nucleus. The NE outflow cone is inclined into the plane of the image, and the SW cone is inclined above the plane.}
	\label{fig:outflow_map}
\end{figure*}

\begin{figure}
	\centering
	\includegraphics[width=0.49\textwidth]{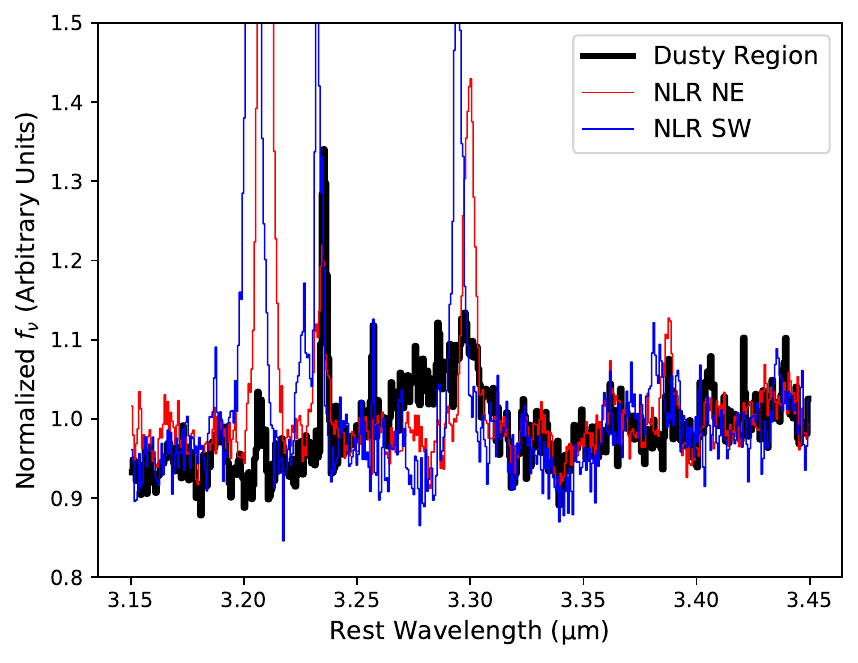}
	\caption{
	Spectra in the ${\sim}3.3\,\mathrm{\mu m}$ window from the apertures used in Figure~\ref{fig:outflow_spectra}. All three spectra are normalized at $3.4\,\mathrm{\mu m}$. The spectra from the NLR apertures do not exhibit a PAH feature, though the level of noise in the continuum is similar to the dusty region. Instead, the NLR spectra have broadened $\mathrm{Pf\,\delta}$ lines at $3.3\,\mathrm{\mu m}$.
	}
	\label{fig:overlayed_specs}
\end{figure}

In Figure~\ref{fig:ArVI_map}, we show the integrated intensity of the [ArVI] line, which traces the NLR. In comparison to the PAH map in Figure~\ref{fig:PAH_map}, this map shows the striking natural separation between the NLR and the PAH-emitting regions, which form two essentially perpendicular distributions. Both of these maps are shown together in Figure~\ref{fig:outflow_map} to better illustrate this separation. These results have implications for the impact of an AGN on PAH grain survival, which are discussed in Section~\ref{sec:survival}. 

Here, we consider one other possible explanation for the lack of PAH detections within the NLR: perhaps the PAH emission is diluted past the point of detection by a much stronger ${\sim}3.3\,\mathrm{\mu m}$ continuum here. In order to gain confidence that we would have detected PAH within the NLR if it were present, we show the ${\sim}3.3\,\mathrm{\mu m}$ spectrum of the dusty region aperture plotted with that of both representative NLR apertures in Figure~\ref{fig:overlayed_specs}, each having been normalized at $3.4\,\mathrm{\mu m}$ (normalization factors are 106.55, 149.37, and 112.76 $\mathrm{MJy\,sr^{-1}}$ for the NE, SW, and Dusty regions, respectively). Indeed, no PAH feature is visible in the outflow spectra, though there is non-continuum emission corresponding to broadened $\mathrm{Pf\,\delta}$ lines at $3.3\,\mathrm{\mu m}$. If the PAH feature were obfuscated by a more significant continuum, the continuum within the outflow regions would be detected with a better SNR than in the dusty region. However, as the level of noise appears to be similar as to the dusty region, we conclude that PAH emission is not present within these representative NLR regions.

\subsection{$H_2$ Emission}

\begin{figure}
\centering
\includegraphics[width=0.49\textwidth]{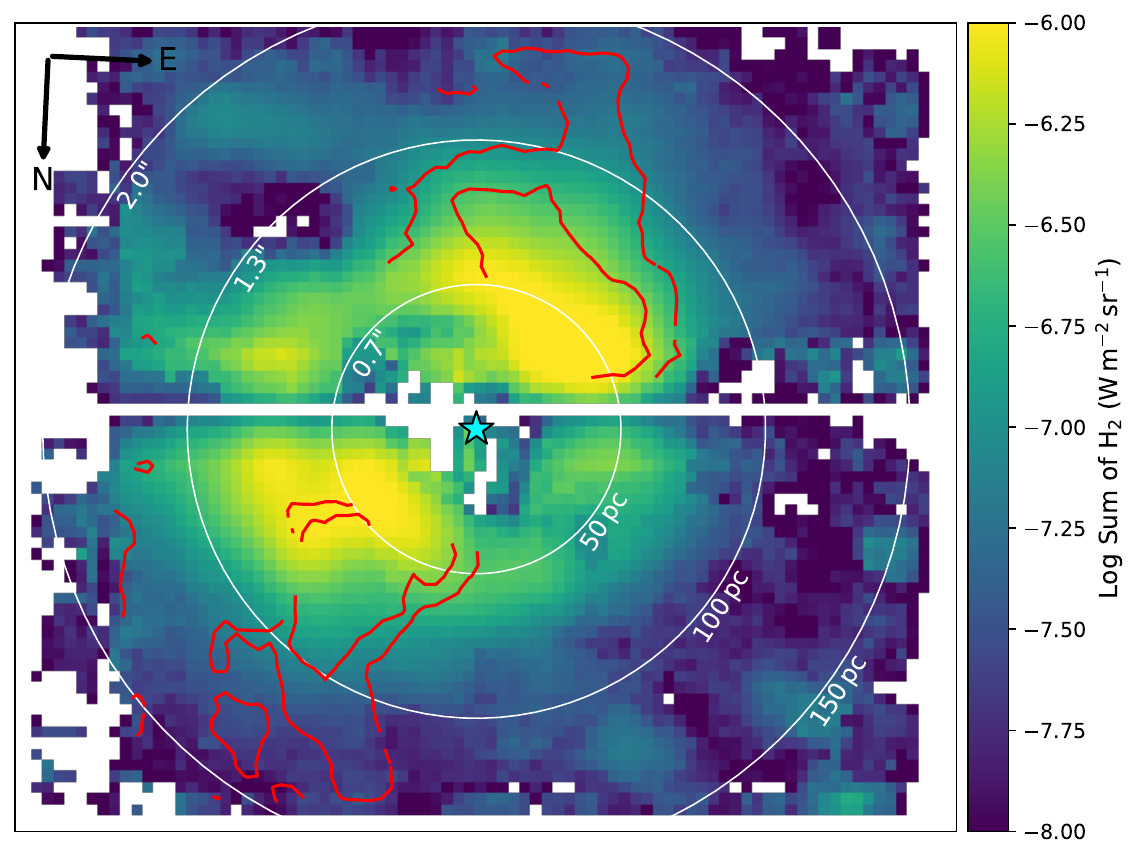}
\caption{
	Map of the total emission of all detected $\mathrm{H_2}$ lines: $\mathrm{H_2\,S(8)}$, $\mathrm{H_2\,S(9)}$, $\mathrm{H_2\,S(13)}$, and $\mathrm{H_2\,O(5)}$. The cyan star and concentric circles are the same as in Figures~\ref{fig:PAH_map} and \ref{fig:ArVI_map}. The red contours show the mean and one standard deviation above the mean of PAH $3.3\,\mathrm{\mu m}$ intensity. Note that the brightest $\mathrm{H_2}$ emission is roughly co-spatial with the PAH $3.3\,\mathrm{\mu m}$ emission.
}
\label{fig:H2_maps}
\end{figure}

Four lines associated with rotational or rovibrational modes of $\mathrm{H_2}$ molecules are detected: $\mathrm{H_2\,1{-}0\,\,O(5)}$, $\mathrm{H_2\,0{-}0\,S(8)}$, $\mathrm{H_2\,0{-}0\,S(9)}$ and $\mathrm{H_2\,0{-}0\,S(13)}$. In Figure~\ref{fig:H2_maps}, we show the spatial distribution of the summed emission of these lines, which arise from very warm $\mathrm{H_2}$ gas; the detected $\mathrm{H_2}$ lines have $E_{\mathrm{upper}}/k$ of 6956, 8677, 10263, and 17458\,K for the list above, respectively. Though some relatively faint $\mathrm{H_2}$ is found in the NLR, the primary $\mathrm{H_2}$-emitting region is generally co-spatial with the PAH-emitting region perpendicular to the NLR, and it is limited to the central ${\sim}100\,\mathrm{pc}$. This spatial distribution is in good agreement with previous NIR $\mathrm{H_2}$ mapping for this region by \cite{storchi-bergmann_2009} and \cite{iserlohe_2013}, where the relative absence of $\mathrm{H_2}$ in the NLR was attributed to molecular dissociation by the AGN. Compared with the high-velocity outflows present in the NLR, \cite{iserlohe_2013} found that the velocity of the $\mathrm{H_2}$ gas is consistent with the normal rotation of material near the nucleus. Similarly, we do not find large velocity offsets in the $\mathrm{H_2}$ gas.

\section{Discussion} \label{sec:discussion}

\subsection{Survival and Destruction of Small PAH Grains} \label{sec:survival}

In the context of the nuclear engine orientation and the location of the NLR, the spatial distribution of PAH $3.3\,\mathrm{\mu m}$ emission provides insight into how small PAH grains can survive or be destroyed in this environment. As we show, small grains are able to survive in even at small radii the region perpendicular to the AGN orientation. On the other hand, any $3.3\,\mathrm{\mu m}$-emitting PAHs that may have been present in the NLR are gone. If PAHs are able to survive in the NLR, it is possible that they have been swept away by the $\gtrsim600\,\mathrm{km\,s^{-1}}$ winds launched by the AGN; recently, dusty outflows from AGN have been detected from velocity shifts in PAH features \citep{donnan_2026b, donnan_2026a}. However, the kinds of extreme conditions found within the NLR are expected to severely limit the lifetime of PAHs, especially the smallest grains carrying the $3.3\,\mathrm{\mu m}$ feature, making it likely that these grains have been destroyed there. Below, we discuss in greater detail two of these PAH destruction pathways: photodestruction, and destruction by sputtering. 

\subsubsection{AGN-driven Photodestruction}
Photodestruction of PAHs by hard AGN radiation, whether from photons in the extreme ultraviolet (EUV) \citep{aitken_roche_1985, desert_1988} or X-ray \citep{voit_1992, siebenmorgen_2004, monfredini_2019}, is perhaps the most commonly suggested mechanism contributing to PAH destruction in galaxies hosting an AGN. The smallest PAHs, which are the primary carriers of the $3.3\,\mathrm{\mu m}$ feature, are thought to be the most susceptible to this type of grain destruction owing to the fact that they have the smallest heat capacities. In NGC\,4151, the geometry of the nucleus and that of the PAH-emitting region is consistent with this; no PAH emission is seen within the NLR bicone above and below the nucleus, and we showed in Section~\ref{sec:results_distros} that we would have detected PAHs if they were there. This region is where grains are exposed to the bare EUV and X-ray radiation from the AGN.

Small PAHs are able to survive in the areas perpendicular to the NLR bicone axis. This implies that if PAHs are destroyed by hard photons in the NLR, the grains surviving outside it must exist in the shadow of some material that shields them. The circumnuclear torus that surrounds an AGN is a commonly suggested mechanism for shielding PAHs \citep[e.g.,][]{esquej_2014, jensen_2017, alonso-herrero_2020}. As the plane of the torus is likely oriented along the direction of the dusty regions (e.g., Figure~\ref{fig:outflow_map}), the existence of abundant warm $\mathrm{H_2}$ gas near the PAHs, but not in the NLR, is consistent with the picture that there is some unresolved structure (i.e., the torus) that protects material along those directions from a significant amount of the direct AGN radiation. The intervening ISM where the $\mathrm{H_2}$ emission is detected may also play a role in shielding the PAHs by pre-processing the radiation from the AGN. Indeed, in Section~\ref{sec:excitation}, we argue that the $\mathrm{H_2}$ emission, which on large scales is co-spatial with the PAHs, is primarily powered by X-ray interactions with the gas.

\subsubsection{Grain Destruction by Sputtering and Shocks} 
PAH molecules can also be destroyed by sputtering. Thermal sputtering occurs when dust grains are exposed to gas of such high temperature that the thermal collisions with particles in the gas strips component atoms off of the grain surface \citep[e.g.,][]{micelotta_sputter}. Similarly, kinetic sputtering (i.e., destruction by shocks) causes mass loss for grains over time, but due to collisions caused by gas turbulence rather than thermal collisions \citep[e.g.,][]{micelotta_shock}. The rate of surface erosion is approximately independent of the grain size for both processes. Therefore, these processes more efficiently destroy smaller grains (which are the primary carriers of the $3.3\,\mathrm{\mu m}$ feature), as these have less initial mass. AGN-driven shocks have been linked to selective small PAH grain destruction as AGN with the highest $\mathrm{H_{2}/PAH}$ ratios tend to also have the lowest short-to-long wavelength PAH ratios in Spitzer data \citep{diamond-stanic_2010, zhang_2022}, which is indicative of a population lacking smaller grains. Though we observe bright $\mathrm{H_2}$ lines in the circumnuclear environment of NGC\,4151, JWST resolves the $\mathrm{H_2}$, showing it lies mostly outside the NLR (and the radio jet), likely due to photodissociation. 

The NLR surrounding the nucleus in NGC\,4151 is composed of hot, X-ray emitting, highly ionized gas. Using Chandra X-ray spectroscopy, \cite{ogle_2000} determined that the NLR gas has a temperature ${\sim}10^{7}$\,K, which limits PAH grain survival to around 1000 years \citep{micelotta_sputter}. At the same time, AGN are thought to produce shocks from interactions between a radio jet and the surrounding ISM; NGC\,4151 is known to have such a radio jet \citep{mundell_2003}, and \cite{storchi-bergmann_2009} argued that the presence of [Fe\,II] lines along and around the jet indicates that it has liberated iron from larger grains via shocks. Recently, it has been suggested that shocks are the dominant mechanism for PAH destruction in lower-luminosity AGN, whereas a more luminous nucleus like that of NGC\,4151 destroys PAHs primarily through photodestruction \citep{zhang_2026a, zhang_2026b}. Sputtering in the NLR plasma is therefore at least a plausible contributor to PAH grain destruction within the NLR. 

\subsection{What is Powering PAH Emission?} \label{sec:excitation}

\begin{figure}
	\centering
	\includegraphics[width=0.49\textwidth]{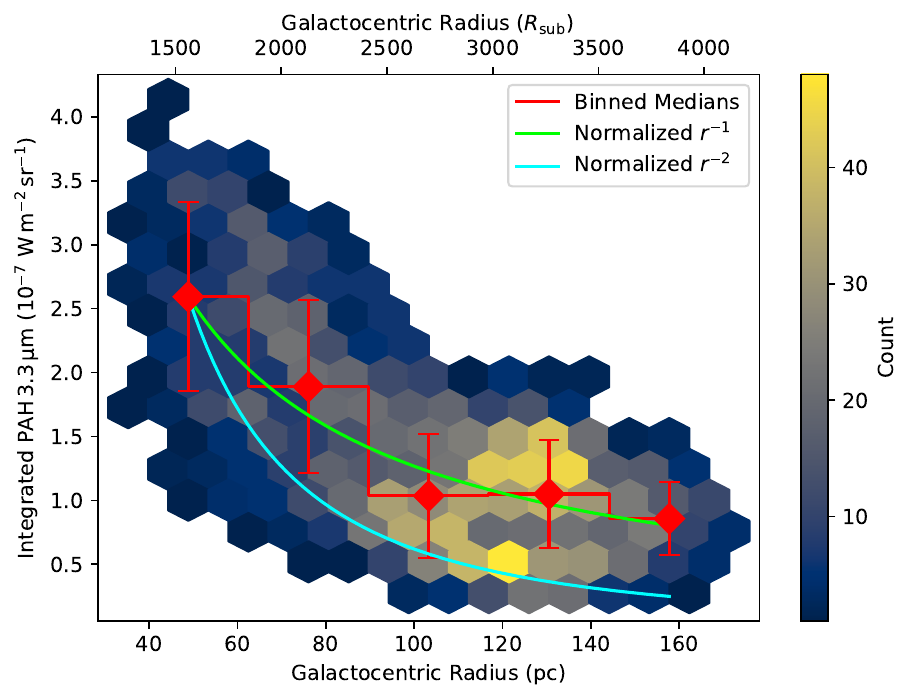}
	\caption{
    In this two-dimensional histogram, the PAH surface flux for each pixel of the PAH map in Figure~\ref{fig:PAH_map} is plotted against its radius from the nucleus. The top axis shows the radius in terms of the dust sublimation radius, $R_{\mathrm{sub}}=0.5\,\mathrm{mas}=0.04\,\mathrm{pc}$ \citep{kishimoto_2022}. The red diamonds show the medians of five bins, while the red stepped line shows the width of each bin and the red errors bars show the standard deviation of a bin. For reference, we show two types of radial profile shapes, both normalized to the intensity of the innermost bin: $r^{-1}$ in green, and $r^{-2}$ in cyan. PAH intensity increases as distance to the nucleus decreases, suggesting that the AGN supplies a portion of the photons for PAH excitation.
	}
	\label{fig:PAH_profile}
\end{figure}

\subsubsection{Evidence for AGN-driven PAH Excitation}

PAH excitation is typically thought to be spurred by radiation from stars, so we consider this case first when it comes to the source of energy for PAH emission. The NIRSpec FOV targets the central bulge of NGC\,4151, which contains primarily evolved stars emitting a softer ISRF compared with star-forming regions. Such an ISRF is known to be sufficient for PAH excitation \citep[for example, in the M31 bulge,][]{groves_2012}, but redder ISRFs like these are the least effective for heating PAH grains enough to produce substantial $3.3\,\mathrm{\mu m}$ emission \citep{draine_21, richie_2026}. On the other hand, bright PAH emission is traditionally associated with recent star formation \citep[e.g.,][]{peeters_2004}, owing to the higher UV absorption cross-sections of small grains compared to other wavelengths. We are not aware of any support in the literature for recent star formation taking place within the central ${\sim}200\,\mathrm{pc}$ of NGC\,4151, and \cite{storchi-bergmann_2009} argued that the observed $\mathrm{H_2\ \lambda2.2477/\lambda2.1218}$ here is not consistent with expected ratios if the $\mathrm{H_2}$ excitation is driven by photons from nearby star formation. Thus, it is unlikely that small PAHs here are being excited primarily by starlight.

An AGN may be able to supply photons for PAH excitation where PAHs are able to survive. Some studies have found that, although galaxies hosting an AGN show overall lower short-to-long wavelength PAH ratios, these ratios can turn up slightly in the nucleus \citep{lai_2022, garcia-bernete_2022b, lai_2023, donnelly_2024}. As such elevated ratios are associated with relatively hotter grains, we suggest that this could indicate that PAHs are heated by an AGN that emits a harder radiation field compared with the surrounding stellar radiation field. Indeed, it was shown by \cite{jensen_2017} using ground-based, high angular resolution MIR spectra that the PAH $11.3\,\mathrm{\mu m}$ surface brightness increases towards the nuclear sources in their sample of AGN hosts. In their sample, there is a nearly universal slope for the PAH surface brightness increasing towards the nucleus within $r<5000\,R_{\mathrm{sub}}$, where $R_{\mathrm{sub}}$ is the dust sublimation radius \footnote{When comparing with other AGN-hosts, it is useful to consider the relevant spatial scales in terms of the dust sublimation radius, $R_{\mathrm{sub}}$, so that distances are inherently scaled to the AGN luminosity.}, whereas at greater distances extended star formation dominates PAH heating. 

If the AGN is the dominant driver of PAH excitation in the circumnuclear environment of NGC\,4151, the PAH intensity radial profile should strictly decrease with distance from the nucleus. In Figure~\ref{fig:PAH_profile}, we show this profile, both in terms of parsecs and $R_{\mathrm{sub}}$; for NGC\,4151, we adopt $R_{\mathrm{sub}}=0.5\,\mathrm{mas}=0.04\,\mathrm{pc}$, which is the upper limit reported using long-baseline NIR interferometry \citep{kishimoto_2022}. The NIRSpec FOV is entirely contained within the $r<5000\,R_{\mathrm{sub}}$ AGN-dominated PAH heating zone that would be expected from the \cite{jensen_2017} sample. Figure~\ref{fig:PAH_profile} shows the clear radial dependence as PAH emission rises towards the nucleus. Though deviations from this shape are to be expected due to spatial structure in the distribution of PAH grains themselves, the surface brightness profile in Figure~\ref{fig:PAH_profile} is remarkably consistent with a profile shape from heating by a compact central source; normalized $r^{-1}$ and $r^{-2}$ profiles are given for reference. Thus, considering that grains are likely destroyed in the NLR, our results paint a more complex picture regarding the impact of an AGN on PAHs. In order for the AGN to destroy PAHs in some regions while powering them in others, energy transferred from the nucleus to PAH grains must first be processed through the shielding medium. Next, we discuss a possible mechanism for this energy transfer.

\subsubsection{PAH Excitation through $H_2$ Reprocessing}

Copious UV photons are produced at the accretion disk of an AGN, and UV photons are typically responsible for PAH excitation. If the small-scale distribution of material in the $\mathrm{H_2}$ and PAH-emitting region is clumpy, it is possible that this UV radiation could illuminate PAHs at the photodissociation fronts of molecular clouds, analogous to the locations of the brightest PAH emission in Galactic photodissociation regions \citep[e.g.,][]{chown_2024}. However, if the material is more evenly distributed, UV radiation may not penetrate far from the nucleus. The torus may also prevent UV radiation from escaping. On the other hand, X-rays leaking through the torus could penetrate much farther into these clouds in order to impart their energy. $\mathrm{H_2}$ molecules ionized upon X-ray photon absorption eject energetic photoelectrons into the ISM, which heat other $\mathrm{H_2}$ molecules through non-thermal collisions. For the case of the $\mathrm{H_2}$ rotational and rovibrational-emitting region shown in Figure~\ref{fig:H2_maps}, both \cite{storchi-bergmann_2009} and \cite{iserlohe_2013} argued that this process could be the dominant driver of the observed $\mathrm{H_2}$ emission, alongside possible shock heating.

Inside of X-ray-irradiated $\mathrm{H_2}$ clouds, \cite{prasad_1983} and \cite{maloney_1996} showed that there is a significant amount of Lyman-Werner and Lyman-alpha UV photons generated from $\mathrm{H_2}$ molecules and H atoms collisionally excited by X-ray liberated photoelectrons. Few of these photons escape the clouds, with the majority being absorbed by dust and PAHs and re-emitted in the IR. We suggest that this may be the primary mechanism for PAH excitation by energy originating at the AGN in NGC\,4151. 

\subsubsection{How Close do PAHs Get to the Nucleus?}

We detect $3.3\,\mathrm{\mu m}$ PAH emission down to ${\sim}50\,\mathrm{pc}$ from the nucleus in shielded regions outside of the NLR. At the same time, there are no indications in Figures~\ref{fig:PAH_map}~or~\ref{fig:PAH_profile} that the trend of increasing PAH surface brightness towards the nucleus does not continue at smaller radii. Is the PAH cutoff at $50\,\mathrm{pc}$ real, i.e., this is the radius where the shielding material is no longer sufficient to protect PAHs, or are PAHs simply not detectable due to aspects of the data quality? Though there are not obvious residual PSF artifacts at this radius in the $5\times5$ spaxel spectra (such as ``ringing" or saturation), this radius is approximately where the innermost and brightest extended spatial structures of the nuclear PSF appear (see Appendix~\ref{appx:PSFsub}, Figure~\ref{fig:PSF_spatial_comparison}). Thus, the lack of PAH detection here is likely due to overwhelming continuum dilution from the residual nuclear PSF, which may increase in brightness faster than the real PAH emission as spaxels approach the nucleus.

Emission lines are less affected by continuum dilution than are PAHs, as these lines are narrower and have a higher peak-to-continuum ratio. In Figure~\ref{fig:H2_maps}, we show $\mathrm{H_2}$ detections considerably farther in, down to ${\sim}25\,\mathrm{pc}$ from the nucleus. Given the roles that we suggest $\mathrm{H_2}$ plays in PAH survival and excitation in this environment, and that PAH surface brightness appears to continue to increase towards the nucleus, we speculate that PAHs may exist farther in than we have detected. Our detected spatial distribution of $\mathrm{H_2}$ emission, including the inner cutoff at ${\sim}25\,\mathrm{pc}$, agrees with the distribution seen using higher resolution, ground-based facilities \citep[e.g.,][]{storchi-bergmann_2009, iserlohe_2013}. Thus, we have likely resolved the minimum radius where $\mathrm{H_2}$ can survive. If PAH grains survive at these smaller radii, it remains to be seen whether there is a radius inside of which PAH emission declines as grains are destroyed by the increasing radiation intensity. However, if PAHs can survive down to radii where UV photons can penetrate the molecular material, these may be the places where PAH emission is brightest due to direct excitation by photons originating from the AGN.

\section{Summary} \label{sec:summary}

The radiative and kinetic feedback from an AGN has long been thought to destroy interstellar PAHs, and perhaps preferentially destroy the smallest of these grains. The details of these destructive mechanisms are unclear, as are the spatial scales where this destruction is effective. In this work, we use new NIRSpec/IFU spectroscopy to shed light on these questions by observing spatially-resolved PAH $3.3\,\mathrm{\mu m}$ emission, which is emitted primarily by the smallest and most fragile grains, within the central ${\sim}200\,\mathrm{pc}$ of the luminous Seyfert NGC\,4151. Our results paint a more complicated picture of the impact of an AGN on small PAH grains than previously expected, with the AGN likely playing a role in both PAH destruction as well as excitation. These are our major findings:

\begin{itemize}
	
	\item We detect the PAH $3.3\,\mathrm{\mu m}$ feature down to 50\,pc from the nucleus, showing that small PAH grains can survive near this AGN, contrary to expectations. PAH detection at some smaller radii is likely limited by data quality, rather than an actual lack of PAH grains.

	\item PAH luminosity increases towards the nucleus in a way that is consistent with heating by a compact central source. We suggest that this is evidence for AGN-driven PAH excitation by internal Lyman-Werner and Lyman-alpha UV photons, produced when X-ray-liberated photoelectrons collisionally excite H and $\mathrm{H_2}$.
	
	\item PAHs are not detected in the bi-conical NLR where they would be exposed to more direct radiation from the nucleus. Instead, PAHs are only detected in the region perpendicular to the orientation of the nuclear engine. This may be due to shielding by the circumnuclear dust torus and intervening ISM, which are aligned with the direction of the PAH detections.
	
\end{itemize}

The circumnuclear environment of NGC\,4151 is an excellent laboratory for studying the impact of an AGN on PAHs, given its proximity and the juxtaposition of NLR and dusty regions. Future work should quantify PAH emission at longer wavelengths using JWST/MIRI spectroscopy, and PAH interband ratios will yield greater insight into PAH size and charge distributions that may help discern which feedback mechanisms are most at play. Just as the nucleus of NGC\,4151 has historically served as an archetype leading to a broader understanding of all AGN, this environment will continue to teach us about how an AGN is an important part of the PAH life cycle in galaxies.

\section*{Acknowledgments} 

This work is based [in part] on observations made with the NASA/ESA/CSA James Webb Space Telescope. The data were obtained from the Mikulski Archive for Space Telescopes at the Space Telescope Science Institute, which is operated by the Association of Universities for Research in Astronomy, Inc., under NASA contract NAS5-03127 for JWST. These observations are associated with program \#8660. Support for US investigators in program \#8660 was provided by NASA through a grant from the Space Telescope Science Institute, which is operated by the Association of Universities for Research in Astronomy, Inc., under NASA contract NAS5-03127. J.-D.T.S. acknowledges visiting support from the Alexander von Humboldt Foundation and the MPIA, which supported a formative pre-JWST workshop at Schloß Hirschhorn, where this research direction took shape.

\appendix 
\restartappendixnumbering

\section{Removing the Nuclear PSF Contamination} \label{appx:PSFsub}

The initial reduced NIRSpec cube for the center of NGC\,4151 is significantly contaminated by a point spread function artifact from the unresolved nucleus. We remove this artifact to reveal the underlying structures. Here, we describe our process for producing an empirical, wavelength-dependent PSF model, and how we used this model to remove the nuclear PSF contamination from each pixel in the NGC\,4151 cube.

\subsection{Empirical PSF Model} 

To constrain the spatial distribution of the contamination from the nucleus, we first construct a model of the wavelength-dependent PSF shape by combining well-sampled NIRSpec IFU observations of stars, i.e., true point sources. Two calibrator stars were observed as part of JWST program 03399 (Co-PIs Perrin and Ruffio) to constrain the NIRSpec IFU PSF. After acquiring the \texttt{uncal} files from MAST, we reduced the observations for both of these stars into spectral cubes using the same settings as we did for NGC\,4151, including the \texttt{adaptive\_trace\_model} option. Critically, all three cubes were produced using the \texttt{ifualign} coordinate system so that the rotation of the PSF on the cube is independent of the telescope roll angle. The stars were observed using a 16 step dither pattern, such that the PSF spatial distribution is better sampled compared with the PSF from the 4 step dither pattern used for NGC\,4151. The cube for the brighter star, HD~163466 (hereafter, ``bright star"), is saturated in the core and presents the same horizontal bar artifact that we see in the cube for NGC\,4151. However, the SNR is high for spaxels at the ``wings" of the PSF. On the other hand, the SNR is high for spaxels at the core of the cube for the ``faint star" (2MASS~J17571324) and there is no saturation, but the SNR at the wings is comparatively poor. Thus, we combine these observations to obtain an empirical PSF model that takes advantage of the strengths of both. Two operations must be done to accomplish this: matching the PSF centroid positions for both stars at each wavelength, and matching their flux scales. 

First, we fit the $x$ and $y$ spaxel positions of the PSF centroids as a function of wavelength using 2-dimensional Gaussians for cube slices $>2806$, for which the bar artifact is no longer visible for the bright star. Finding that the centroids drift with wavelength at the sub-pixel level, we fit lines to the $x$ and $y$ positions as a function of wavelength, and we extrapolate the centroid positions for slices $\leq2806$. We then use these fits to reposition each slice of the faint star cube so that it is spatially aligned with the bright star. As both stars are Spitzer and JWST photometric standards, their magnitudes and NIR spectra are well-known. We obtained the \texttt{CALSPEC} calibrated model spectra for both stars from the Space Telescope Science Institute\footnote{\url{https://www.stsci.edu/hst/instrumentation/reference-data-for-calibration-and-tools/astronomical-catalogs/calspec}} \citep{CALSPEC1, CALSPEC2, CALSPEC3, CALSPEC4}, and we used the ratio of these model spectra to scale the aligned faint star cube to match the bright star. Finally, we replaced any saturated and/or bar-masked pixels in the bright star cube with pixels from the aligned and scaled faint star cube, producing the final empirical PSF model.

\subsection{PSF Decomposition Method} \label{sec:PSF_method}

The spectrum $I_{xy}(\lambda)$ corresponding to spaxel $xy$ in the NGC\,4151 cube is a linear combination of the real, independent extended emission, $E_{xy}(\lambda)$, and the contamination from the nuclear PSF, $N_{xy}(\lambda)=C_{xy}n(\lambda)$. Here, $n(\lambda)$ is the normalized shape of the unresolved spectrum of the nucleus, and $C_{xy}$ is its scaling factor. For spaxel $xy$, the relative contributions of $E(\lambda)$ and $N(\lambda)$ to $I(\lambda)$ is unknown, and initially, the shape of $n(\lambda)$ is unknown. As the same wavelength-dependent PSF artifact affects all spaxels (albeit in different amounts), it follows that $n(\lambda)$ is the same everywhere. Additionally, the magnitude of $C_{xy}$ for one spaxel compared with another can be obtained from the empirical PSF model.

We decompose $I_{A}(\lambda)$ for spaxel A by comparing it with $I_{B}(\lambda)$ for an \emph{adjacent} spaxel B. The difference between spaxel spectra is

\begin{equation}
	\begin{aligned}
		I_{B}(\lambda) - I_{A}(\lambda) = (E_{B}(\lambda) - E_{A}(\lambda)) + (C_{B}-C_{A})n(\lambda)\ .
	\end{aligned}
\end{equation}

\noindent From the PSF model spatially aligned to the NGC\,4151 centroids, we obtain $R_{BA} \equiv C_{B}/C_{A}$, which gives

\begin{equation}
	\begin{aligned}
		I_{B}(\lambda) - I_{A}(\lambda) = (E_{B}(\lambda) - E_{A}(\lambda)) + (R_{BA}-1)C_{A}n(\lambda)
	\end{aligned}
\end{equation}

\noindent by substitution. We make the assumption that $E(\lambda)$ comes mostly from smooth spatial structures, and that very sharp real spatial structures with high contrast between adjacent pixels are rare. Thus, $E(\lambda)$ for two adjacent spaxels is assumed to be similar, i.e., $E_{B}(\lambda)-E_{A}(\lambda)=\epsilon(\lambda)$. If $\epsilon(\lambda)$ is negligible compared to the differential spectrum of the nuclear contamination for the pair of A and B spaxels, then rearranging gives

\begin{equation}
	\begin{aligned}
		C_{A}n(\lambda) = N_{A}(\lambda) \approx  \frac{ I_{B}(\lambda) - I_{A}(\lambda) }{ R_{BA}-1 } \ .
	\end{aligned}
\end{equation}

The $n(\lambda)$ obtained this way for a given comparison between spaxels tends to be noisy. Likewise, both $n(\lambda)$ and each $C_{xy}$ derived from one spaxel comparison can accumulate errors if the above assumptions do not hold; if the adjacent spaxels fall on sharp spatial features, and/or there is low contrast between spaxels in the PSF shape (i.e., if $R$ is near unity). However, the noise and errors are mitigated substantially by averaging over the results derived from the ensemble of comparisons between each spaxel and all eight of its neighbors (including diagonals). Finally, we scale the aligned empirical PSF model using the average $N(\lambda)$, and we subtract this from the NGC\,4151 cube. 

\subsection{PSF Subtraction Quality Checks} \label{sec:PSF_quality}

\begin{figure}
	\centering
	\includegraphics[width=0.99\textwidth]{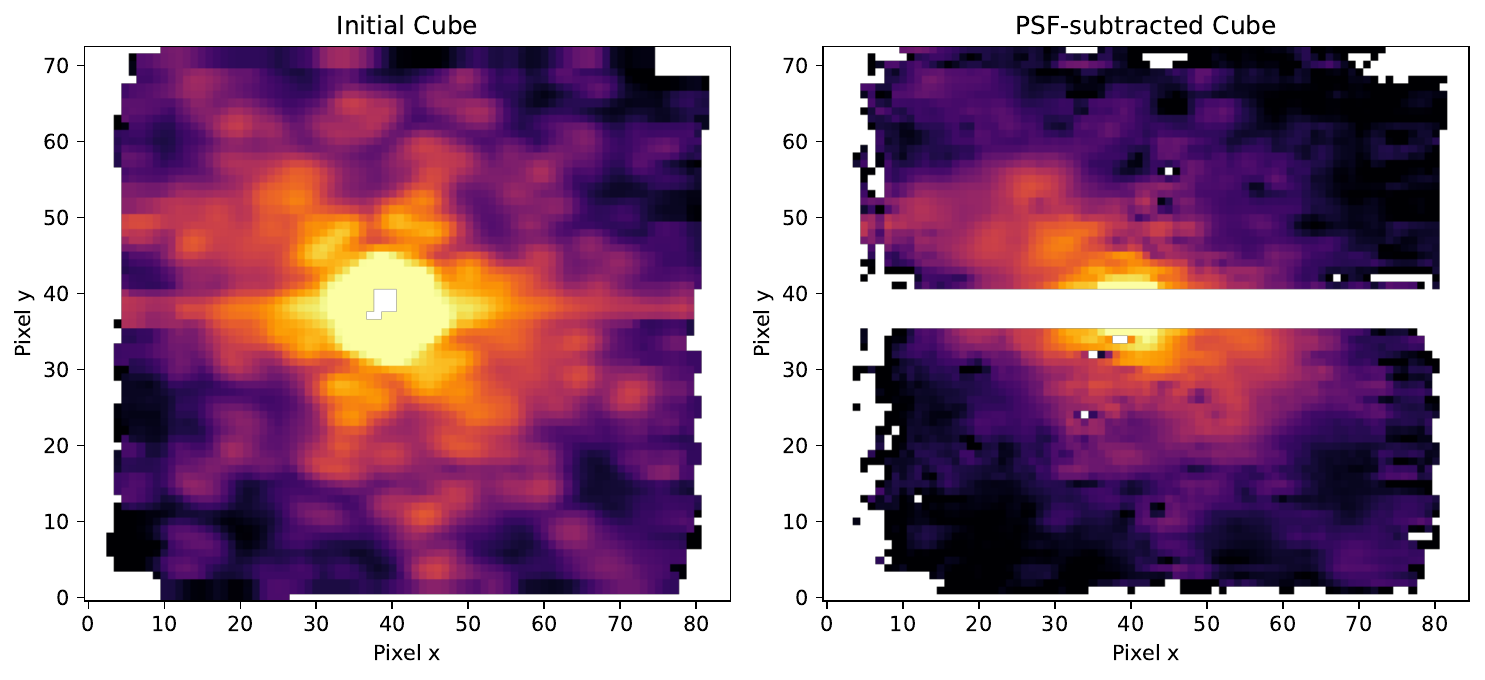}
	\caption{
	Comparison between the initial NGC\,4151 cube (left) and the PSF-subtracted cube (right), where each image is the moment-zero sum between slice indices 2505--2535, capturing the [ArVI] line across its full observed range of velocity offsets. Both panels have been stretched to a consistent logarithmic scale. The horizontal bar artifact is visible in the left panel, but it has been intentionally masked in the right panel. These images are displayed in the IFU pixel coordinate system.
	}
	\label{fig:PSF_spatial_comparison}
\end{figure}

Here, we provide some examples of the spatial and spectral structure of the PSF subtraction to show its quality. In Figure~\ref{fig:PSF_spatial_comparison}, we show a comparison between the initial NGC\,4151 cube and its PSF-subtracted counterpart, summed across the wavelengths of the observed [ArVI] line. This range of cube slices contains the brightest emission line in the NLR, so it represents the best chance for any coherent spatial distribution to be visible in the initial PSF-dominated cube. The PSF artifact still completely dominates the spatial structure of the initial cube, and the NLR (which was traced by the [ArVI] line emission) cannot be made out. However, after the PSF-subtraction, the shape of the NLR can easily be seen.

\begin{figure}
	\centering
	\includegraphics[width=0.99\textwidth]{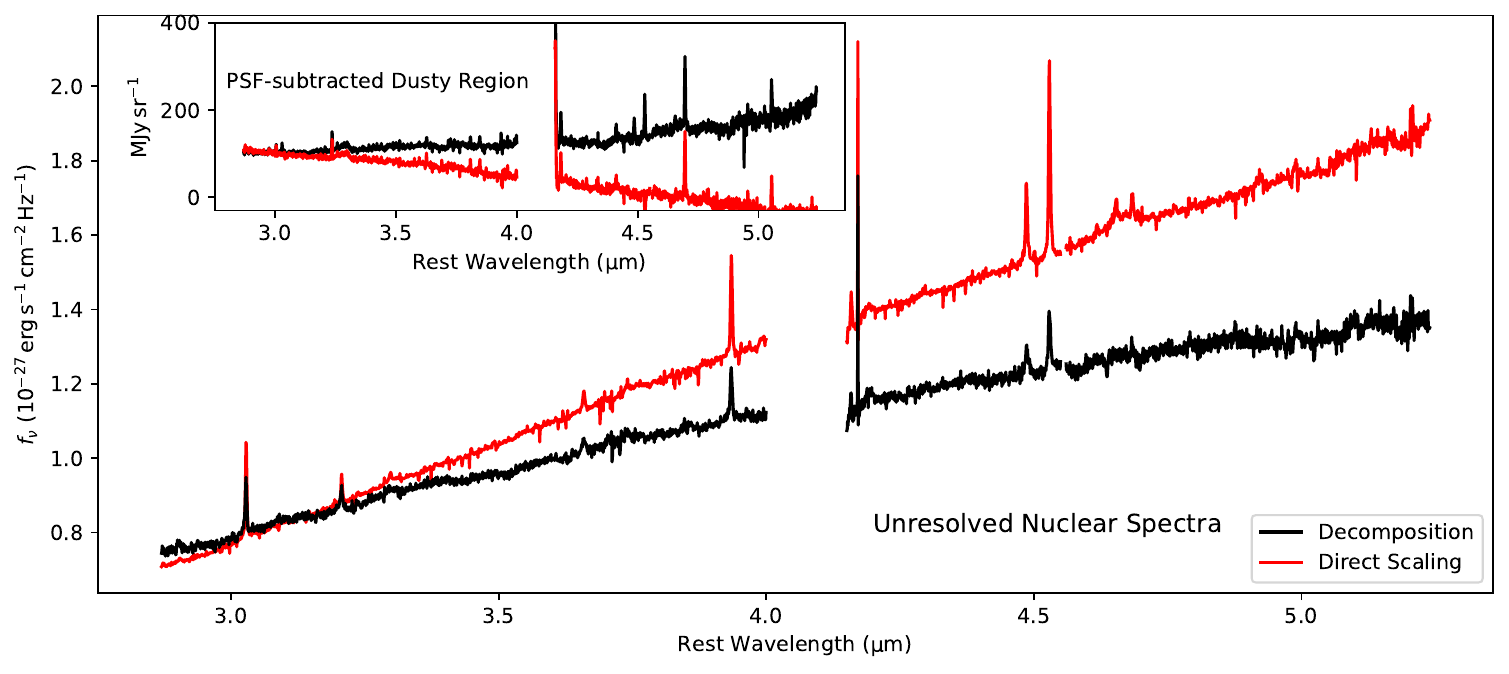}
	\caption{
		The main plot shows a comparison between the unresolved nuclear spectra recovered from two different methods for the PSF subtraction: the method described in Section~\ref{sec:PSF_method} (black, ``Decomposition"), and a direct scaling method described in Section~\ref{sec:PSF_quality} (red). The Decomposition method was used throughout this work. The inset plot shows spectra from the Dusty region resulting from either PSF subtraction method. 
	}
	\label{fig:PSF_spectral_comparison}
\end{figure}

Perhaps the most straightforward way to use the empirical PSF model would be to simply scale the model such that the intensities of the center-most spaxels (which are the most PSF-dominated) are matched with those in the NGC\,4151 cube, then subtract. This is not practical for our case because the center most spaxels within the NGC\,4151 cube are saturated or are affected by the bar artifact. A compromise could be to scale by the spaxels with the greatest intensity that are not affected by saturation or by the bar. This direct scaling method could serve as a reasonable comparison for the performance of the above method. Thus, for comparison, we applied the direct scaling method by locating the spaxels in the NGC\,4151 cube that are above the $98^{\mathrm{th}}$ percentile in median brightness across all wavelengths. This excludes saturated/bar pixels and includes bright pixels near the centroid. Then, we scaled the aligned PSF model such that the sum of these pixel values in both cubes are equal. Finally, we subtract the scaled PSF model.

In Figure~\ref{fig:PSF_spectral_comparison}, we show the unresolved nuclear spectrum as recovered by both the decomposition used in this work (Section~\ref{sec:PSF_method}), and the direct scaling method described above. The spectra from both methods have similar flux values, which shows that our method is isolating an appropriate fraction of spaxel spectra as being due to nuclear contamination. We expect the direct scaling method to have generally higher flux because it incorporates some degree of extended emission as well. There are two main differences: the $f_{\nu}$ slope is higher for the direct scaling method, and the direct scaling method has brighter emission lines. A possible reason for this higher slope is due to the growing FWHM of the PSF at longer wavelengths; more flux from the center is distributed to the pixels used for the scaling, which surround the center but do not include it. The PSF-subtracted dusty region spectra using either method is shown in the inset of Figure~\ref{fig:PSF_spectral_comparison}. Here, while the continuum for the decomposition method is relatively flatter, and it appears only to start to rise at longer wavelengths, the direct scaling method produces a negatively sloped continuum that approaches zero surface brightness. In certain other regions of the cube using the direct scaling subtraction, the continuum has negative values. Finally, we take the emission line flux for the direct scaling method to also be an overestimate as some portion of this comes from the non-nuclear emission. Indeed, there are negative lines from over subtraction seen in the direct scaling subtracted cube, but these are not present with the decomposition method. For these reasons, we find that the decomposition method described in Section~\ref{sec:PSF_method} does a better job than the alternative direct scaling method.

\newpage
\bibliography{references}{}
\bibliographystyle{aasjournalv7}

\end{document}